\documentclass[a4paper]{JHEP3}
 
\usepackage{amsmath,amsfonts,a4,graphicx}
\usepackage{thophys}

\numberwithin{equation}{section}
\def\beq{\begin{equation}}
\def\eeq{\end{equation}}
\def\bea{\begin{eqnarray}}
\def\eea{\end{eqnarray}}

\def\gsim{\mathrel{\raisebox{-.6ex}{$\stackrel{\textstyle>}{\sim}$}}}
\DeclareMathOperator{\tr}{tr}

\newcommand{\spinl}[3]{\Braket{{#1}|#2|#3}}

\preprint{KA-TP-16-2009, SFB/CPP-09-96}

\title{NLO QCD corrections to graviton production at hadron colliders}

\author{Stefan Karg, Michael Kr\"amer\\
  Institut f\"ur Theoretische Physik, RWTH Aachen University, D-52056
  Aachen, Germany}
\author{Qiang Li, Dieter Zeppenfeld\\
  Institut f\"ur Theoretische Physik, Karlsruhe Institute of
  Technology, D-76128 Karlsruhe, Germany}

\abstract {Models with large extra dimensions predict the existence of
  Kaluza-Klein graviton resonances.  We compute the next-to-leading
  order QCD corrections to graviton plus jet hadro-production, which
  is an important channel for graviton searches at the Tevatron and
  the LHC.  The QCD corrections are sizable and lead to a significant
  reduction of the scale dependence. We present numerical results for
  cross sections and distributions, and discuss the uncertainty from
  parton distribution functions and the ultraviolet sensitivity of the
  theoretical prediction.}

\date{\Date}


\begin{document}

% --------------------------------------------------------------------

\section{Introduction}\label{sec:1}

The search for new physics at the TeV-scale is one of the major tasks
for current and future high-energy physics experiments.  Models with
extra space dimensions and TeV-scale gravity address the problem of
the large hierarchy between the electroweak and Planck scales, and
predict exciting signatures of new physics that can be probed at
colliders~\cite{Giudice:2008zza}.

In the $D=4+\delta$ dimensional model proposed by Arkani-Hamed,
Dimopoulos and Dvali (ADD)~\cite{ADD}, the standard model (SM)
particles are constrained to a $3+1$ dimensional brane, while gravity
can propagate in a $4+\delta$ dimensional space-time.  For simplicity,
the additional $\delta$-dimensional space is assumed to be a torus
with common compactification radius $R$. In such a model, the
4-dimensional effective Planck scale $M_{\rm P}$ is related to the
fundamental scale $M_{\rm S}$ by~\cite{ADD}:
\begin{eqnarray}\label{scale}
M^2_{\rm P}=8\pi R^\delta M^{\delta+2}_{\rm S}\,.
\end{eqnarray}
For a large compactification radius $R$ it is thus possible that the
fundamental scale is near the weak scale, $M_{\rm S} \sim {\rm TeV}$.
In the ADD model, deviations from the standard Newton law of gravity
are predicted at distances around $R\approx 0.83 \times
10^{-16+30/\delta}\,{\rm mm} \times (2.4\,{\rm TeV}/ M_{\rm
  S})^{1+2/\delta}$. Current terrestrial test of gravity exclude $R
\ge 37 (44)\, \mu$m for $\delta = 2(1)$~\cite{Kapner:2006si}, which,
using Eq.~(\ref{scale}), translates into $M_{\rm S} \ge 3.6$~TeV for
$\delta = 2$. Further constraints have been derived from astrophysics
and cosmology, in particular for $\delta < 4$, but they can be evaded
in specific models~\cite{Kaloper:2000jb} and do not lessen the
importance of collider searches for extra dimensions.

The $D=4+\delta$ dimensional graviton corresponds to a tower of
massive Kaluza-Klein (KK) modes in 4 dimensions. The interaction of
these spin-2 KK gravitons with SM matter can be described by an
effective theory~\cite{feynrules2, Mirabelli:1998rt, feynrules1} with
the Lagrangian
\begin{eqnarray}\label{IL}
 {\cal L}_{\rm int} = - \frac{1}{{\overline M}_{\rm P}}
\sum_{\vec{n}} G^{(\vec{n})}_{\mu \nu} T^{\mu
\nu}\,,
\end{eqnarray}
where the massive gravitons are labeled by a $\delta$-dimensional
vector of integers, $\vec{n}=(n_1,n_2,..,n_\delta)$, ${\overline
  M}_{\rm P}=M_{\rm P}/\sqrt{8 \pi}\sim 2.4\times 10^{18}$~GeV is the
reduced 4-dimensional Planck scale, and $T_{\mu \nu}$ is the
energy-momentum tensor of the SM fields. The Feynman rules that follow
from Eq.~(\ref{IL}) can be found in
Refs.~\cite{feynrules2,feynrules1}.  The individual KK resonances have
masses equal to $m_{(\vec{n})} = |\vec{n}|/R$ and thus the mass gap
between neighboring modes $\Delta m=R^{-1}$ is small for $\delta$ not
too large. Quantitatively one finds $\Delta m \approx$ 20~keV, 7~MeV
and 0.1~GeV for $M_{\rm S} =1$~TeV and $\delta=$4, 6 and 8,
respectively~\cite{feynrules2}. The discrete mass spectrum can thus be
approximated by a continuum with a density of states ${\rm d}N =
\rho(m) {\rm d}m$~\cite{feynrules2,Mirabelli:1998rt}, where
\begin{eqnarray}\label{rho}
\rho(m)= S_{\delta-1}\frac{{\overline
M}_{\rm P}^2}{M_{\rm S}^{2+\delta}}m^{\delta-1},\, \,\, \, \rm{and}\, \,
S_{\delta-1}=\frac{2\pi^{\delta/2}}{\Gamma(\delta/2)}.
\end{eqnarray}
Inclusive collider cross sections, where one sums over all accessible
KK modes, are obtained from a convolution of the cross section for an
individual KK mode of mass $m$, ${\rm d}\sigma_m$, with the mass
density function $\rho(m)$~(\ref{rho}), ${\rm d}\sigma/{\rm d}m =
\rho(m){\rm d}\sigma_m$.  Although each individual graviton couples to
SM matter with only gravitational strength $\propto 1/{\overline
  M}_{\rm P}$, see Eq.~(\ref{IL}), and thus ${\rm d}\sigma_m \propto
1/{\overline M}_{\rm P}^2$, inclusive collider processes are enhanced
by the enormous number of accessible KK states $\propto {\overline
  M}_{\rm P}^2$ (\ref{rho}). The factors ${\overline M}_{\rm P}^2$
cancel in ${\rm d}\sigma/{\rm d}m = \rho(m){\rm d}\sigma_m$, leaving
an overall suppression of only $M_{\rm S}^{-2-\delta}$. If the
fundamental scale $M_{\rm S}$ is near the TeV-scale, graviton
production can thus be probed at present and future high-energy
colliders.

Both virtual graviton exchange between SM particles and real graviton
emission provide viable signatures of large extra dimensions at
colliders. Since the coupling of gravitons with matter is suppressed
$\propto 1/{\overline M}_{\rm P}$, direct graviton production gives
rise to missing energy signals. Searches for graviton production have
been performed in the processes $e^+e^- \to \gamma(Z) + E^{\rm miss}$
at LEP and $p\bar{p}\to \gamma({\rm jet}) + p_T^{\rm miss}$ at the
Tevatron. The combined LEP limits~\cite{LEP} read $M_{\rm S}>$ 1.60,
1.20, 0.94, 0.77, 0.66~TeV, for $\delta=$ 2,$\cdots$,6 respectively,
while Tevatron searches exclude $M_{\rm S} >$ 1.40, 1.15, 1.04, 0.98,
0.94~TeV, for $\delta=$ 2,$\cdots$,6
respectively~\cite{Aaltonen:2008hh, Abazov:2008kp}. Searches for the
process $pp\to {\rm jet} + p_T^{\rm miss}$ at the LHC will be able to
extend the sensitivity to the fundamental scale $M_{\rm S}$ into the
multi-TeV region~\cite{LHCG, Wu:2008hw, CMSI, CMSII}.

Current analyses of graviton production at hadron colliders are based
on LO cross sections, which are subject to large theoretical
uncertainties from the choice of renormalization and factorization
scales. In this paper we present the first calculation of the NLO QCD
corrections to graviton production in the process $pp/p\bar{p}\to {\rm
  jet} + G$ at hadron colliders,
results for the QCD corrections in the photon channel have been presented in   
\cite{Gao:2009pn}.
The NLO cross sections lead to
significantly more accurate theoretical signal predictions and thereby
more accurate constraints on $M_{\rm S}$ or, in the case of discovery,
will allow to probe the model parameters. The NLO calculation also
enables us to properly study the relative importance of multi-jet
final states, which has been addressed in a recent calculation of the
(tree-level) graviton plus di-jet cross section~\cite{G2j}. We note
that NLO QCD corrections to the hadro-production of lepton and boson
pairs in models with large extra dimensions have been presented in a
series of recent papers~\cite{Mathews:2004xp,ravindran-add}.

The remainder of this paper is structured as follows: In
section~\ref{sec:2} we present some details of the NLO QCD cross
section calculation for $pp/p\bar{p} \to {\rm jet} + G$. Numerical
results for the Tevatron and the LHC are presented in
section~\ref{Numericalresults}.  We conclude in
section~\ref{sec:summary}. More details of the calculation and
selected formulae are presented in the appendix.

% --------------------------------------------------------------------

\section{Calculation}
\label{sec:2} 

The LO cross section for graviton plus jet production receives
contributions from the partonic processes
\begin{equation}\label{eq:xshat}
  q\bar{q}\rightarrow gG,\quad qg\rightarrow qG \quad {\rm and}\quad
  gg\rightarrow gG\,.
\end{equation}
The relevant Feynman diagrams are depicted in Fig.~\ref{fig:LO diags}.
The LO partonic cross sections of the processes (\ref{eq:xshat}) have
first been presented in Ref.~\cite{feynrules2}; the corresponding
helicity amplitudes are discussed in the appendix.

\FIGURE{\label{fig:LO diags}
\includegraphics[width=12cm]{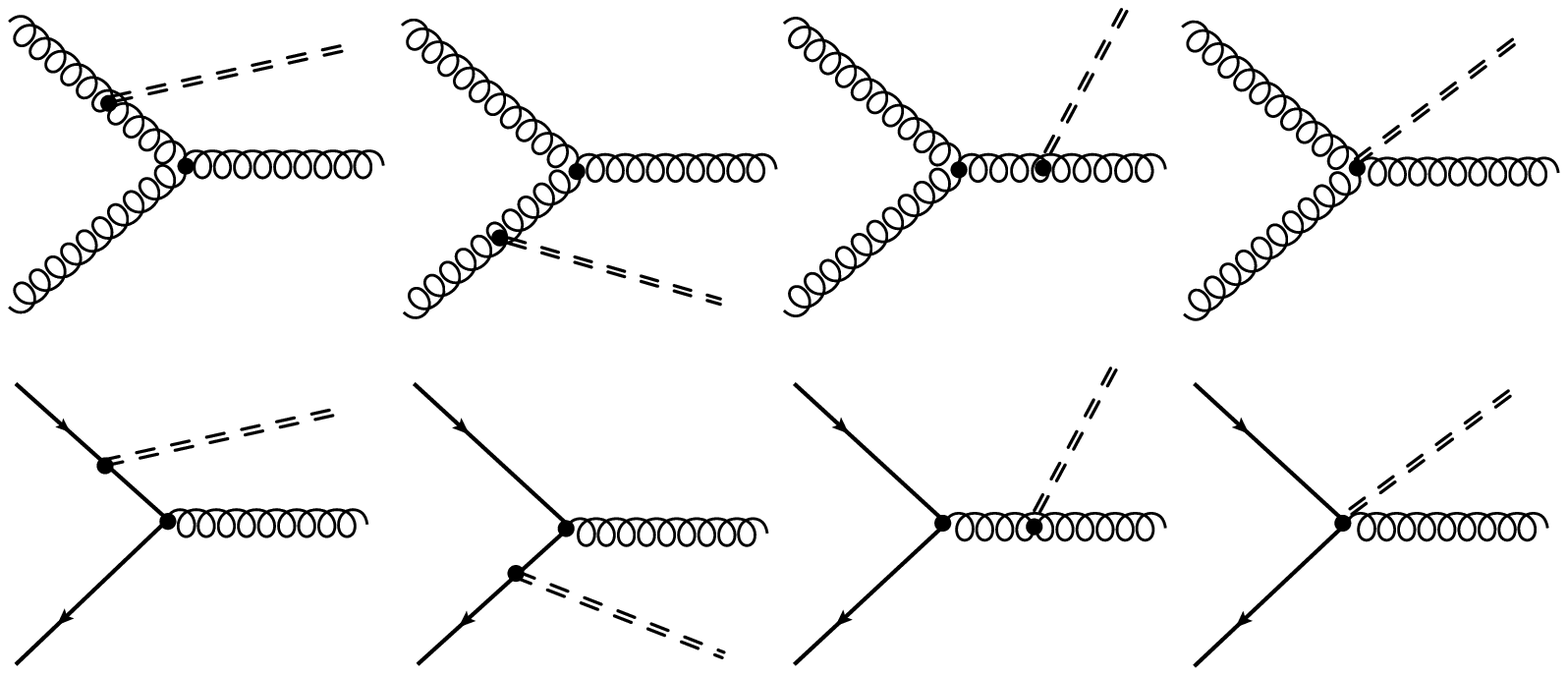}
\caption{The LO Feynman diagrams for $gg, q\bar q\to G+$jet. The $gq$
  channel is related to the $q\bar{q}$ channel by crossing and not
  shown.}}

The NLO cross section consists of virtual corrections, real-emission
contributions and a collinear term, which is a finite remainder of the
factorization of collinear singularities into the parton distribution
functions (PDFs). We use dimensional regularization~\cite{DREG} in
$d=4-2\epsilon$ dimensions to regulate the ultraviolet (UV) and
infrared (IR) divergences, and apply the dipole subtraction
scheme~\cite{CS} to cancel the infrared singularities. The UV
divergences are removed by renormalization of the QCD coupling
$\alpha_{\rm s}$ in the $\overline{\textrm{MS}}$-scheme.

We have performed two independent calculations of the virtual
corrections, described in more detail below, and have checked gauge
invariance and Ward identities arising from general coordinate
invariance, see Ref.~\cite{G2j} for more details. The numerical
implementation of the real-emission contributions is based on {\tt
  MadGraph}~\cite{spin2MG} and {\tt MadDipole}~\cite{MadDipole}.  Some
details of the NLO calculation are provided below.

\subsection{Virtual corrections}
\label{Vir}
The virtual corrections to $pp/p\bar{p} \to {\rm jet} + G$ arise from
the interference of the Born and one-loop amplitudes. Example one-loop
Feynman diagrams are depicted in Fig.~\ref{fig:virt diags}. The
Feynman rules for the graviton interaction with the SM fields can be
found in
Ref.~\cite{feynrules2,feynrules1,Mathews:2004xp}\footnote{Note that
  there are different sign conventions in the definition of the
  covariant derivative, which leads to a sign difference in the
  Feynman rules for 4-point vertices such as $VVVG$.}. For external
gluons we choose the light-cone gauge to avoid introducing external
ghost lines, as in the {\tt HELAS} convention~\cite{helas}. The
internal gluon propagators are evaluated in the 't Hooft-Feynman
gauge, and the unphysical degrees of freedom are canceled by ghost
loops.  The UV poles are removed by
$\overline{\textrm{MS}}$-renormalization of $\alpha_{\rm s}$, while
the IR singularities cancel against those from the real emission
contribution (see Eq.~(\ref{Ipole}) below).

\FIGURE{\label{fig:virt diags}
 \includegraphics[]{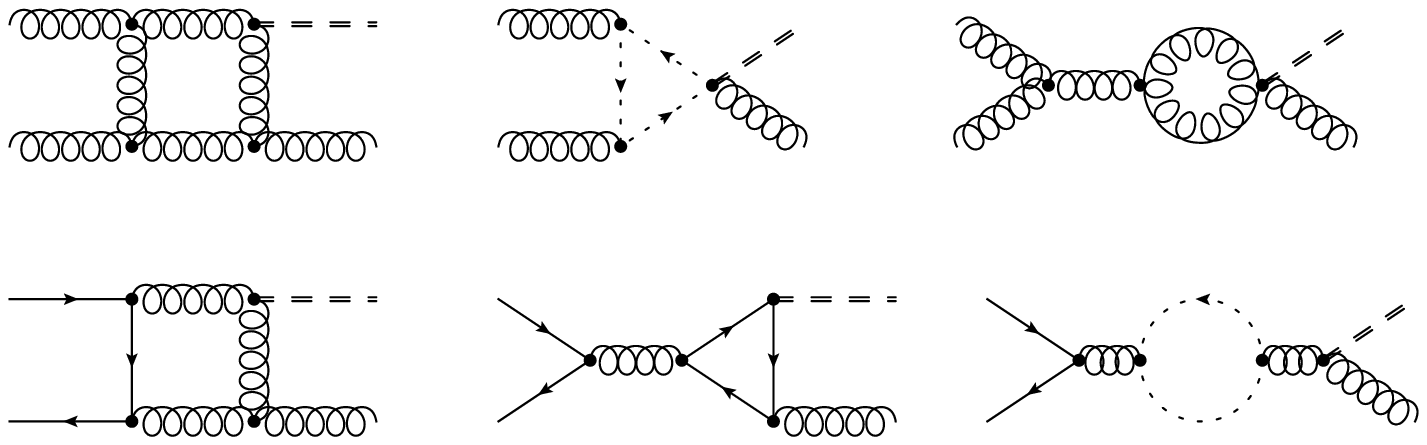}
 \caption{Examples of NLO QCD virtual Feynman diagrams for $pp\to
   G+$jet. The dotted loops represent ghost particles.}}

Two independent calculations are performed for the virtual
corrections. The first calculation is based on the Mathematica package
{\tt FeynCalc}~\cite{FeynCalc}. Because of the Lorentz indices of the
spin-2 graviton, we encounter high-rank tensor integrals, such as
rank-5 4-point functions. Special care is taken to reduce those to
one-loop scalar integrals by an independent Mathematica code,
following the prescription of Ref.~\cite{rank5}.

In the second calculation, the one-loop diagrams are generated with
{\tt QGRAF}~\cite{qgraf} and then projected onto helicity components
and amplitude coefficients with {\tt FORM}~\cite{form}. Details of the
projection are given in the appendix
and can also be found in Ref.~\cite{amplitudesetup}, where the same 
method has been used.
The tensor reduction is
performed according to the {\tt
  GOLEM}~\cite{Binoth:2005ff,Binoth:2006rc} reduction algorithm,
supplemented with additional tensor reduction routines for rank $N+1$
$N$-point tensor integrals 
with $N\leq 3$.
To calculate numerical results we employed the OmniComp-
Dvegas package~\cite{OmniComp}, which facilitates parallelised 
adaptive  Monte Carlo integration and was developed in the 
context of Ref.~\cite{OmniComp_references}.

\subsection{Real-emission contributions}

The real-emission contribution comprises the radiation of a real gluon
or a massless (anti-)quark. Soft and collinear singularities are
isolated using dipole subtraction~\cite{CS}. Collinear emission from
initial state partons is factorized into the parton distribution
functions defined in the $\overline{\textrm{MS}}$-scheme. The
remaining IR and IR/collinear singularities, which cancel those of the
virtual corrections, read in the notation of~\cite{CS}
\begin{eqnarray}\label{Ipole}
  && {\langle I(\epsilon)\rangle }_{gg}=\frac{\alpha_s}{2\pi}\frac{(4\pi\mu^2)^\epsilon}{\Gamma (1-\epsilon)}\bigg\{\frac{3\beta_0}{2\epsilon}+\frac{C_A}{\epsilon^2}[(s)^{-\epsilon}+(-t)^{-\epsilon}+(-u)^{-\epsilon}]\bigg\}|{\cal{M}}_{gg}^{ \rm{B}}|^2, \\
  && {\langle I(\epsilon) \rangle }_{q\bar{q}}=\frac{\alpha_s}{2\pi}\frac{(4\pi\mu^2)^\epsilon}{\Gamma (1-\epsilon)}\bigg\{
  \frac{\beta_0}{2\epsilon}+\frac{3C_F}{2\epsilon}+\frac{C_A}{\epsilon^2}[(-t)^{-\epsilon}+(-u)^{-\epsilon}] \nonumber\\
  &&\hspace{2.cm}+\frac{(-C_A+2C_F)}{\epsilon^2}(s)^{-\epsilon}\bigg\}|{\cal{M}}_{q\bar{q}}^{ \rm{B}}|^2,
\end{eqnarray}
where $s$, $t$, and $u$ are the Mandelstam variables,
$\beta_0=(11C_A-4n_fT_R)/3$ with $n_f=5$, and ${\langle
  I(\epsilon)\rangle }_{gq}$ can be obtained by crossing from
${\langle I(\epsilon)\rangle }_{q\bar{q}}$.

The real emission matrix elements and subtraction terms are generated
with {\tt MadGraph} with Spin-2 particles~\cite{spin2MG} and {\tt
  MadDipole}~\cite{MadDipole}, respectively, and are implemented in a
parton-level Monte Carlo program.

% --------------------------------------------------------------------

\section{Numerical Results}
\label{Numericalresults}

In this section we present NLO cross sections for $pp/p\bar{p} \to
{\rm jet} + G$ at the Tevatron ($\sqrt{S}$ = 1.96 TeV) and the LHC
($\sqrt{S}$ = 14 TeV).  Before we proceed with the numerical results,
we note that the interaction of the KK gravitons with SM matter is
described by an effective theory~\cite{feynrules2, Mirabelli:1998rt,
  feynrules1}, which is valid only for scattering energies
$\sqrt{\hat{s}}$ smaller than the fundamental scale $M_{\rm S}$. While
hadron collider cross sections in principle involve partonic
scatterings with energies $\sqrt{\hat{s}}$ up to the collider energy
$\sqrt{S}$, the rapid decrease of the parton luminosities at large
$\sqrt{\hat{s}}$ suppresses the high-energy region and allows for a
cross section prediction that is not very sensitive to the UV
completion of the effective theory. We will return to this issue at
the end of the section and provide quantitative estimates of the UV
sensitivity of the theoretical prediction.

The numerical results presented below are obtained with $\alpha_{\rm
  s}$ and the parton distribution functions defined in the
$\overline{\textrm{MS}}$-scheme, with five active flavours.  Throughout
our calculation, we employ the 2008 MSTW LO(NLO) PDF~\cite{MSTW2008}
at LO(NLO), with the corresponding value for the strong coupling
$\alpha_{\rm s}$.  Our default choice for the renormalization and
factorization scale is the transverse momentum of the graviton, $\mu =
P_T^G$.

To suppress SM backgrounds in the LHC graviton searches~\cite{LHCG},
we require
\begin{equation}\label{lhcset}
P_T^{\rm miss}> 500~{\rm GeV}\,.
\end{equation}
The jets are defined by the $k_{T}$ algorithm, with the resolution
parameter set to $D=0.6$, and are required to satisfy $|\eta_j|<4.5$
and $P_T^j>50$~GeV.

For the Tevatron predictions we use the same settings as in the recent
CDF study~\cite{Aaltonen:2008hh}, i.e.\
\begin{equation}\label{tevset}
  P_T^{\rm miss}> 120~{\rm GeV}\,,\quad
  P_T^{j}>150~{\rm GeV}\,,\quad {\rm and} \quad |\eta_j|<1\,.
\end{equation}
Also here, jets are defined by the $k_{T}$ algorithm with $D=0.7$, and
are required to satisfy $|\eta_j|<3.6$ and $P_T^j>20$~GeV. A second
jet with $P_T>60$~GeV is vetoed. We will, however, also discuss
results without the jet-veto.

\FIGURE{\label{scaleplot}
\includegraphics[width=14cm]{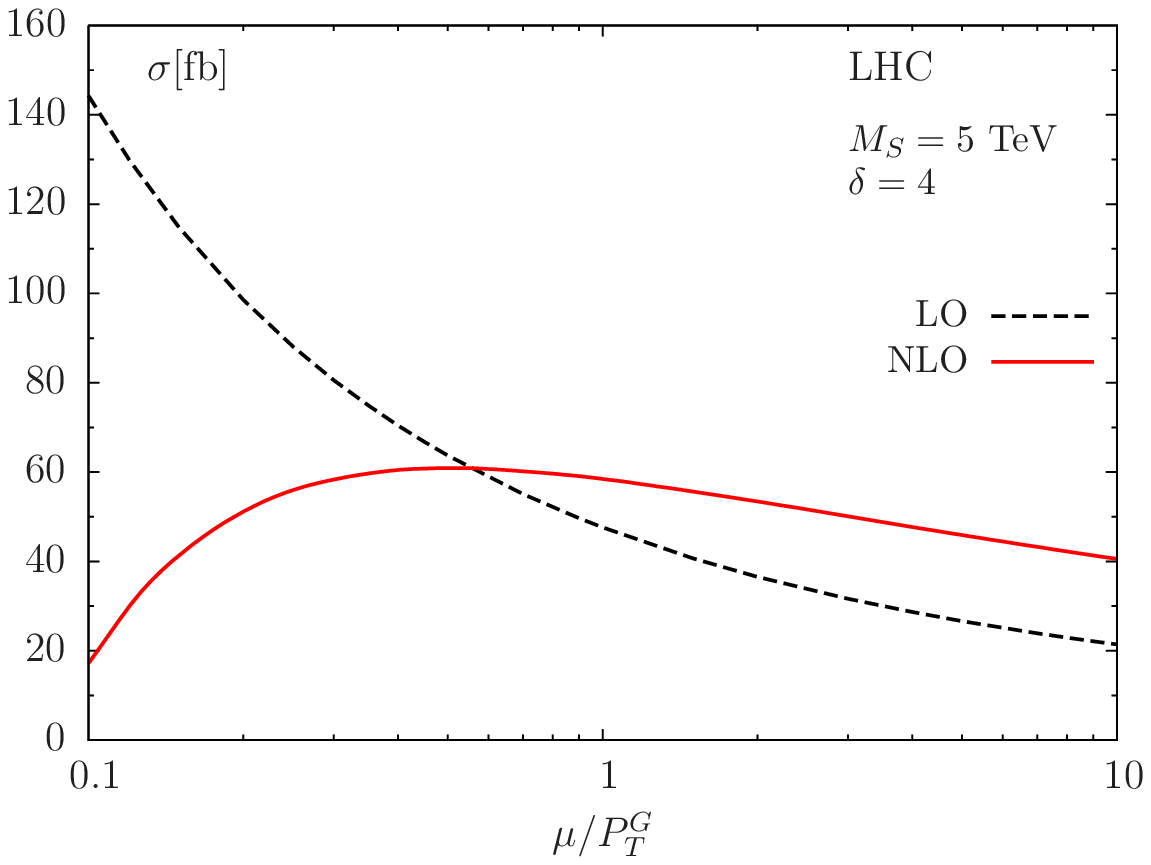}
\includegraphics[width=14cm]{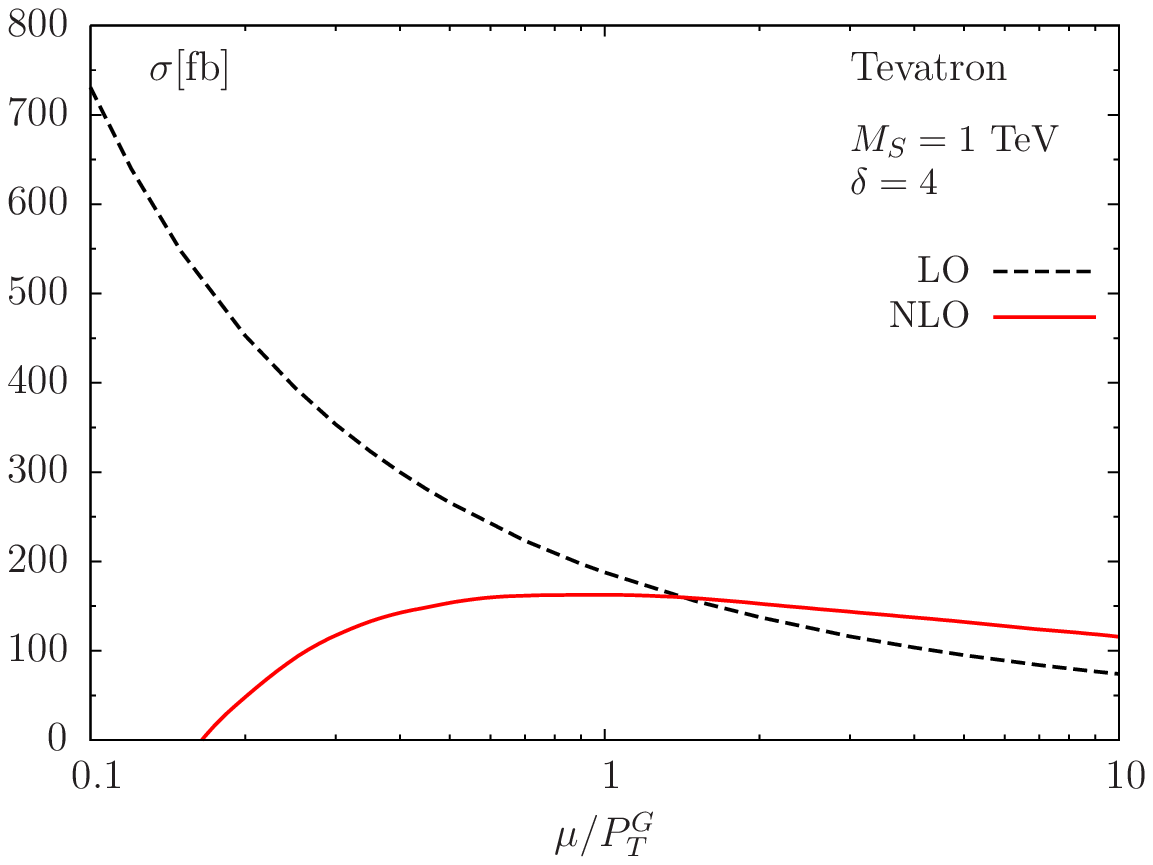}
\caption{Scale variation for the integrated cross section at LHC and
  Tevatron, for a common scale $\mu=\mu_r=\mu_f$. Selection cuts are
  described in the text.}}

We first focus on the scale dependence of the total cross section.
For illustration, we set the model parameters to $\delta=4$, and
$M_{\rm S} = 5$~TeV (LHC) and 1~TeV (Tevatron). 
Note that the cross section scales as $\sigma \propto M_{\rm S}^{-2-\delta}$ 
so that results for other values of $M_{\rm S}$ can be obtained by rescaling our
predictions. 
Unfortunately, it is not possible to determine both $M_{\rm S}$ and
$\delta$ independently from just the shape of the
$P_T^{\mathrm{miss}}$ spectrum~\cite{LHCG}. To resolve
these parameters would need very accurate measurements at
different hadron collider center-of-mass energies; the ratio of graviton
production cross sections at different center-of-mass energies depends
on $\delta$ through the kinematic limit on the graviton mass, while
the dependence on $M_{\rm S}$ cancels. Operating the LHC at 
7~TeV and 14~TeV center-of-mass energies may offer such an 
opportunity.
We impose the
kinematical cuts listed in Eqs.~(\ref{lhcset}) and (\ref{tevset}) for
the LHC and the Tevatron, respectively. The LO and NLO results are
shown in Fig.~\ref{scaleplot} as a function of the renormalization and
factorization scales varied around the central scale $\mu = P_T^G$. We
observe that the scale dependence of the NLO cross section is
significantly smaller than that of the LO cross section, both at the
LHC and at the Tevatron: changing $\mu$ in the range between $P_T^G/2
$ and $2P_T^G $, the LO cross section varies by $\approx 30\%$, while
the scale uncertainty at NLO is less than $\approx 10\%$. We have also
varied both scales independently and find that in all cases the NLO
uncertainty is less than approximately 10\%. At the LHC, the
$K$-factor, $K=\sigma_{\rm NLO}/\sigma_{\rm LO}$, is sizeable and
positive at the central scale $\mu = P_T^G$, increasing the LO cross
section prediction by about $20\%$. At the Tevatron, the QCD
corrections are mild near $\mu=P_T^G$ with $K\approx 1$, but are
essential to reduce the theoretical uncertainty.

\FIGURE{\label{PTplot1}
\includegraphics[width=14.cm]{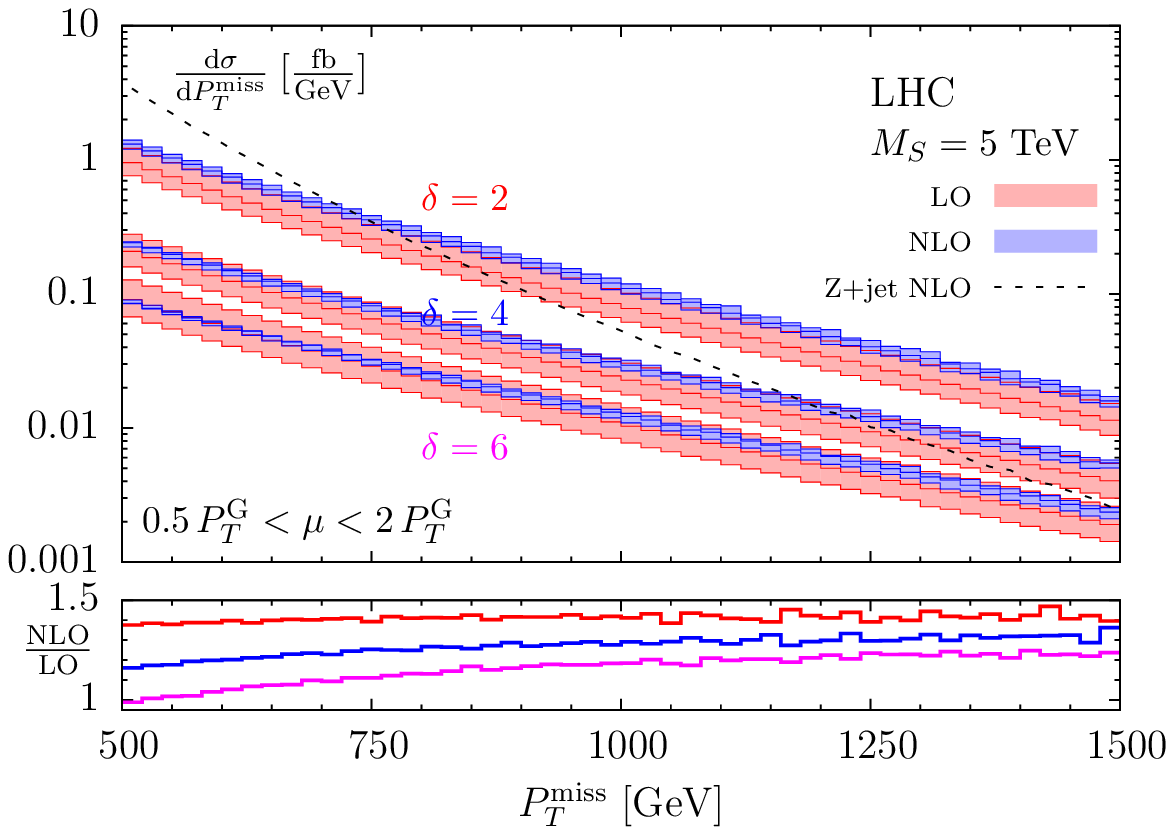}
\caption{$P_T^{\mathrm{miss}}$ distribution for the graviton signal at
  the LHC with scale uncertainty bands ($0.5 \,P_T^{\mathrm{G}} < \mu
  <2\,P_T^{\mathrm{G}}$). Also given is the NLO distribution for the
  dominant $Z\to \nu\bar\nu$ background. The lower part of the plot
  shows $K(P_T)=({\rm d}\sigma_{\rm NLO}/{\rm d}P_T)/({\rm
    d}\sigma_{\rm LO}/{\rm d}P_T)$ for
  $\delta=2,4,6$ (top down).}}

\FIGURE{\label{PTplot2}
\includegraphics[width=14.cm]{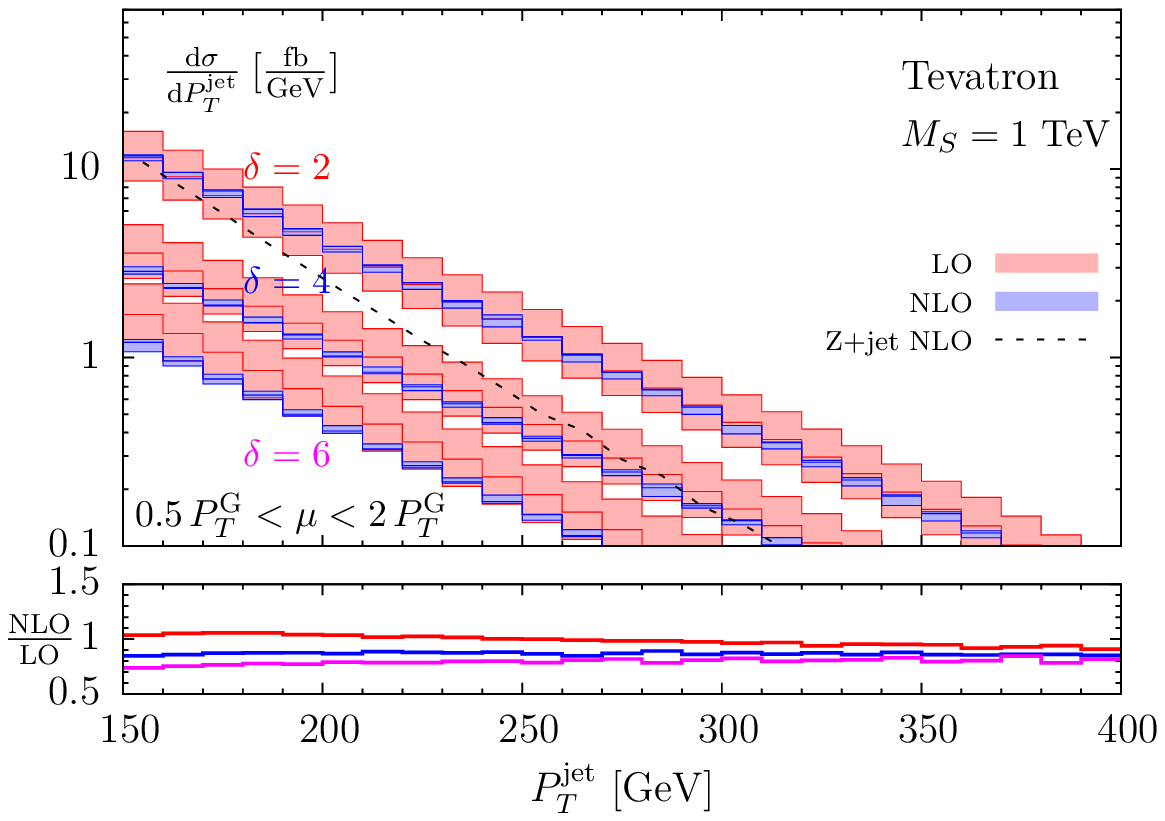}
\caption{Same as Fig.~\ref{PTplot1} but for the $P_T$ distribution of
  the leading jet at the Tevatron. }}

The experimental analyses at the LHC and the Tevatron rely on the
$P_T^{\mathrm{miss}}$ and $P_T^{\mathrm{jet}}$ distributions,
respectively. In Figs.~\ref{PTplot1} and ~\ref{PTplot2} we show the
scale dependence of these distributions, for different choices of the
number of extra dimensions $\delta=2,4,6$.  Current Tevatron 
limits appear to exclude $M_{\rm S} = 1$~TeV for $\delta < 4$. However, 
these analyses are based on leading-order predictions 
and should be refined using the NLO results presented here. 
We thus include numerical predictions for $M_{\rm S} = 1$~TeV 
and $\delta= 2,4,6$ below. We also show the NLO QCD
predictions for the main background $pp\rightarrow
Z(\to\nu\bar{\nu})+$\,jet obtained with {\tt MCFM}~\cite{mcfm}. 
[Note that establishing an excess in graviton plus
jet production at hadron colliders requires an excellent experimental
understanding of the SM background. To precisely estimate the dominant
$pp\rightarrow Z(\to\nu\bar{\nu})+$\,jet background process one can
rely on a calibration sample of the related process $pp\rightarrow
 Z(\to e^+e^-/\mu^+\mu^-)+$\,jet~\cite{ Aaltonen:2008hh,
Abazov:2008kp,LHCG,CMSI}. Furthermore, the signal to background ratio can be 
improved by increasing the $P_T^{\rm miss}$-cut.]
The
bands show the uncertainty of the LO and NLO predictions when varying
the renormalization and factorization scales in the range $P_T^G/2 <
\mu <2\,P_T^G$. The reduction of the scale uncertainty at NLO is
evident.  Figs.~\ref{PTplot1} and \ref{PTplot2} also display the $P_T$
dependence of the $K$-factors, defined as $K(P_T)=({\rm d}\sigma_{\rm
  NLO}/{\rm d}P_T)/({\rm d}\sigma_{\rm LO}/{\rm d}P_T)$. The $K$
factors are sizeable at the LHC (Fig.~\ref{PTplot1}), as noted before,
increasing with decreasing $\delta$.  Furthermore, the $K$-factors
depend on the kinematics and increase with increasing
$P_T^{\mathrm{miss}}$. At the Tevatron, the $K$-factors are in general
near or below one and only mildly depend on the jet transverse
momentum, see Fig.~\ref{PTplot2}.

We have also investigated the uncertainty of the NLO cross section
prediction due to the parton distribution function. Using the MSTW
error PDFs~\cite{MSTW2008}, we find an uncertainty of less than
approximately 15\% for graviton production at the LHC, even for large
$P_T^{\mathrm{miss}} > 1$~TeV. At the Tevatron, the uncertainty is
even smaller and approximately 5\%.

Let us now examine the contribution of the real emission cross section
with two hard jets. In Fig.~\ref{vetoed} we show the ratio of the
cross section where we require two hard jets with $P_T^{j}> P_T^{\rm
  cut}$ and the inclusive cross section, as a function of
$P_T^{\mathrm{miss}}$ and 
the leading
$P_T^{\mathrm{jet}}$ at the LHC and the
Tevatron, respectively.  Results are presented for the minimum
jet-$P_T$ set to $P_T^{\rm cut} = 150$~GeV and 250~GeV for the LHC,
and $P_T^{\rm cut} = 60$~GeV for the Tevatron. Moreover, in
Fig.~\ref{vetoed}, we show results with an alternative choice of
settings (labeled B in the plot) defined as
\begin{equation}\label{eq:scale_B}
\mu_f = {\rm min}(P^j_T ) \quad {\rm and} \quad 
\alpha_s=\sqrt{\alpha_s(P_T^{j_1}) \; \alpha_s(P_T^{j_2}})\,,
\end{equation}
which was used in Ref.~\cite{G2j} for the real emission contribution
with two hard jets. We observe that the fraction of events containing
two jets with $P_T^{j}> 250$~GeV is 20-40\% at the LHC for
$P_T^{\mathrm{miss}}$ above 1~TeV, depending on the choice of input
parameters for the 2-jet contribution. Between 40\% and 70\% of the
events with $P_T^{\mathrm{miss}} > 1$~TeV contain two jets with
$P_T^{j}> 150$~GeV. We note that even for the scale choice B, as given
in Eq.~(\ref{eq:scale_B}), the fraction of di-jet events is smaller
than estimated in Ref.~\cite{G2j}. This is due to the denominator,
i.e.\ the larger inclusive cross section as predicted at NLO.
Nevertheless, also with our new estimate we expect a large fraction of
high $P_T^{\mathrm{miss}}$ events with two or more hard jets at the
LHC.  
While the quantitative estimates given in Ref.~\cite{G2j} are
thus changed due to the impact of the NLO corrections, the qualitative
conclusions remain valid.
At the Tevatron, the contribution of 2-jet events with $P_T^{j}>
60$~GeV is moderate and does not exceed 20\%. The difference between
the results with the two different scale settings, which represent
part of the uncertainty for the (tree-level) 2-jet cross section,
increases with increasing $P_T$, as the difference between our default
choice of scale and $\alpha_{\rm s}$ and the alternative choice
(\ref{eq:scale_B}) becomes larger.

\FIGURE{\label{vetoed}
\includegraphics[width=14cm]{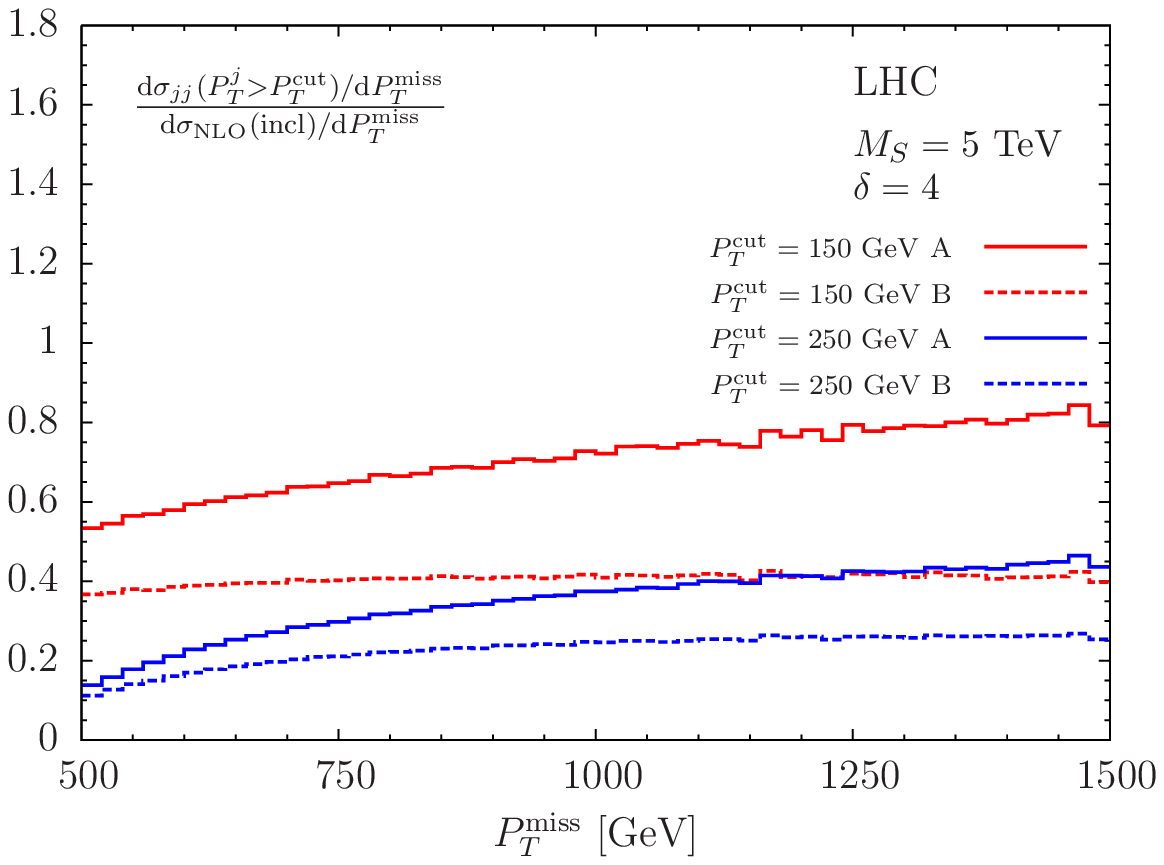}
\includegraphics[width=14cm]{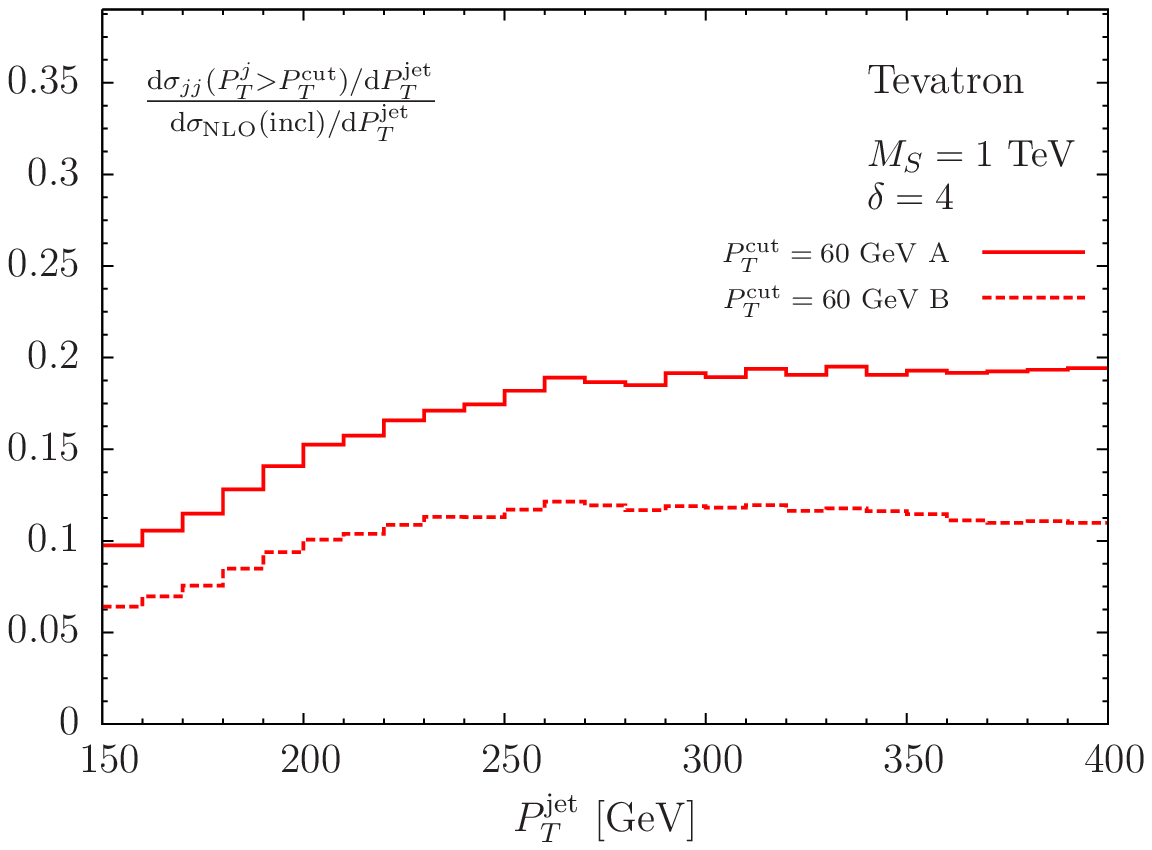}
\caption{Di-jet fraction of graviton plus jets events at the LHC and
  Tevatron. Results are given for the two scale choices $\mu=p_T^G$ and
  Eq.~(\ref{eq:scale_B}), respectively, and minimal
  transverse momentum requirements $P_T^{\mathrm{cut}}$ for the second
  jet.  }}

\FIGURE{\label{truncation1}
\includegraphics[width=14.cm]{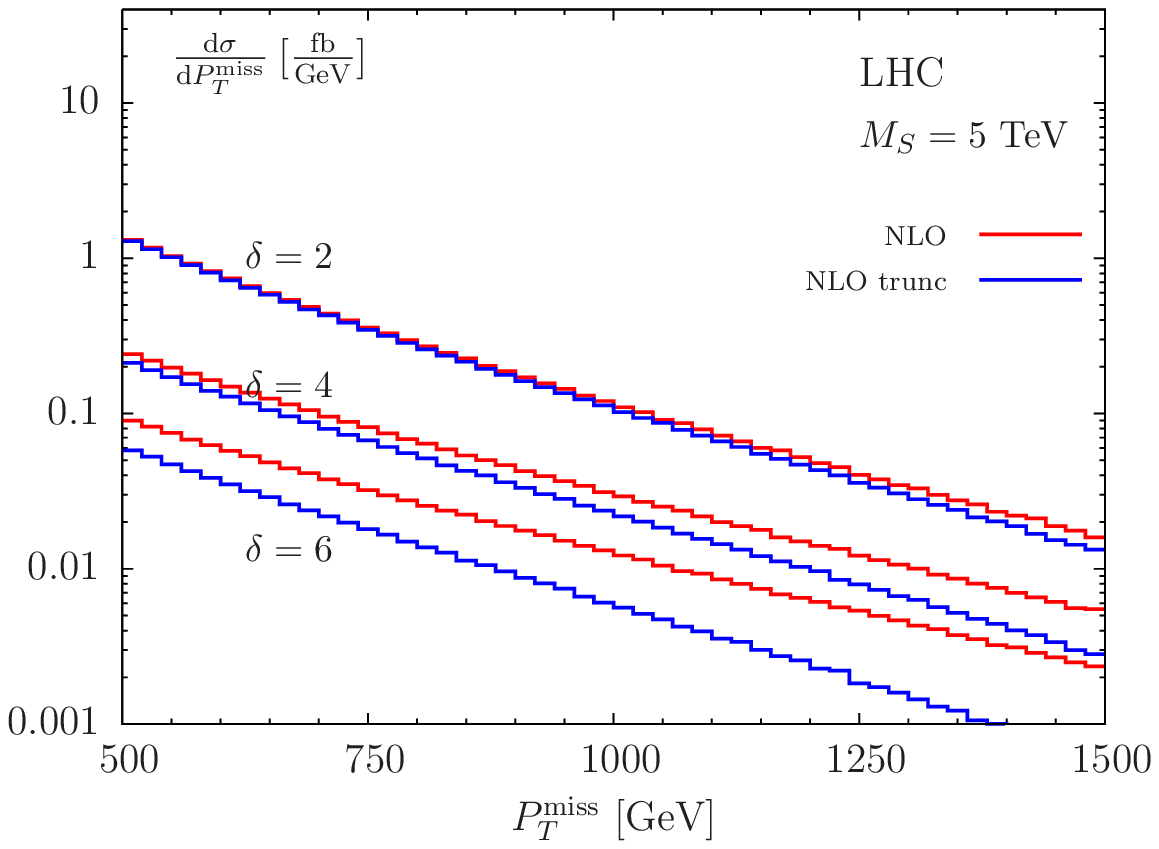}
\caption{Effect of truncation of the partonic cross section above 
$Q_{{\rm truncation}}=M_S$ at the LHC. See text for details.}}

\FIGURE{\label{truncation2}
\includegraphics[width=14.cm]{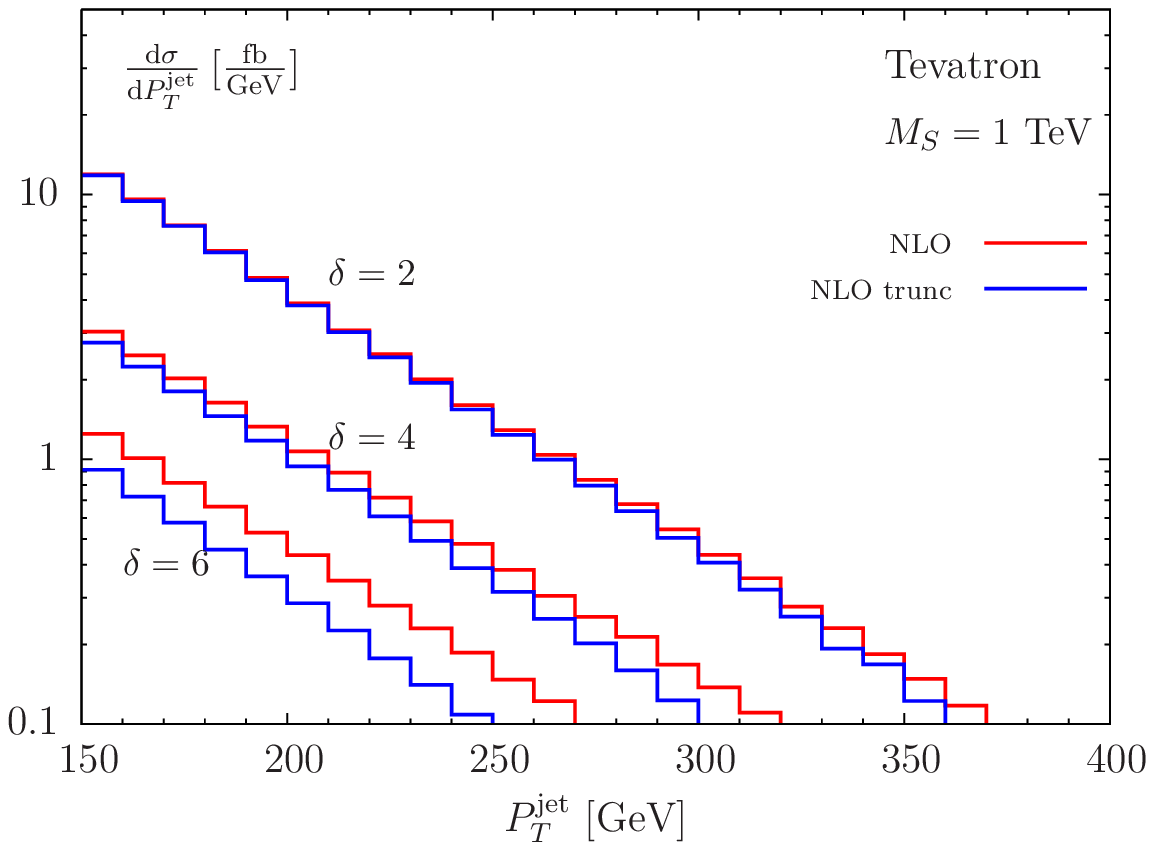}
\caption{Effect of truncation of the partonic cross section above 
$Q_{{\rm truncation}}=M_S$ at the Tevatron. See text for details.}}

As mentioned at the beginning of this section, the results of the
effective field theory calculation are valid only as long as the
scales involved in the hard scattering process do not exceed the
fundamental scale $M_{\rm S}$. To quantify the sensitivity of our
prediction to the unknown UV completion of the theory, we compare our
NLO results with those involving a truncation scheme which sets the
cross section to zero if $Q_{{\rm truncation}} \ge M_{\rm S}$. In the
numerical results presented below, the truncation parameter $Q_{{\rm
    truncation}}$ is taken to be the invariant mass of the missing
momentum and observable jet(s),
\begin{eqnarray}
  Q_{{\rm truncation}} = |P^G + P^{\rm{jet(s)}}|\,.
\end{eqnarray}
This definition is equal to $Q_{{\rm truncation}} = \sqrt{\hat{s}}$ at
LO, but takes into account that the effective partonic energy of the
scattering process can be reduced by collinear initial state radiation
at NLO and thus provides an IR-safe definition of the truncation
parameter. NLO results for the transverse momentum distributions with
and without the hard truncation scheme for $\delta=2,4,6$ at the LHC
and the Tevatron are shown in Figs.~\ref{truncation1} and
\ref{truncation2}, respectively. As expected, the differences between
the two calculations increase with increasing $\delta$, as the average
graviton mass is shifted to larger values.  Also, the differences
become larger as $P_T^{\mathrm{miss}}$ ($P_T^{\mathrm{jet}}$)
increase. For example, for the LHC, at $P_T^{\mathrm{miss}}=
1250$~GeV, the uncertainties for $\delta=2,4,6$ are about $5\%$,
$20\%$, and $50\%$, respectively, while for the Tevatron, at
$P_T^{\mathrm{jet}}= 250$~GeV, the uncertainties for $\delta=2,4,6$
are about $2\%$, $10\%$, and $25\%$. Note that the results with the 
hard truncation do not obey the simple scaling $\sigma \propto 
M_{\rm S}^{-2-\delta}$.

Finally, we comment on the prospects for graviton searches during the initial
phase of the LHC operating at 7~TeV. Even at half the nominal center-of-mass 
energy, the LHC will be able to extend the sensitivity of current Tevatron 
searches to larger values of $M_{\rm S}$~\cite{CMSI, CMSII}. For illustration 
we show in Fig.~\ref{7tev} the NLO $P_T^{\rm miss}$ distribution for $M_{\rm S} = 2$~TeV 
and $\delta = 2,4,6$, together with the dominant SM background. We find 
sizeable signal rates, in particular for $\delta=2$, which exceed the background 
for $P_T^{\rm miss} \gsim 250$~GeV. We re-emphasize that comparing mono-jet 
signatures at 7~TeV with results obtained at higher center-of-mass energies during a 
later stage of LHC operation may allow to disentangle the fundamental parameters 
$M_{\rm S}$ and $\delta$.
\FIGURE{\label{7tev}
\includegraphics[width=14.cm]{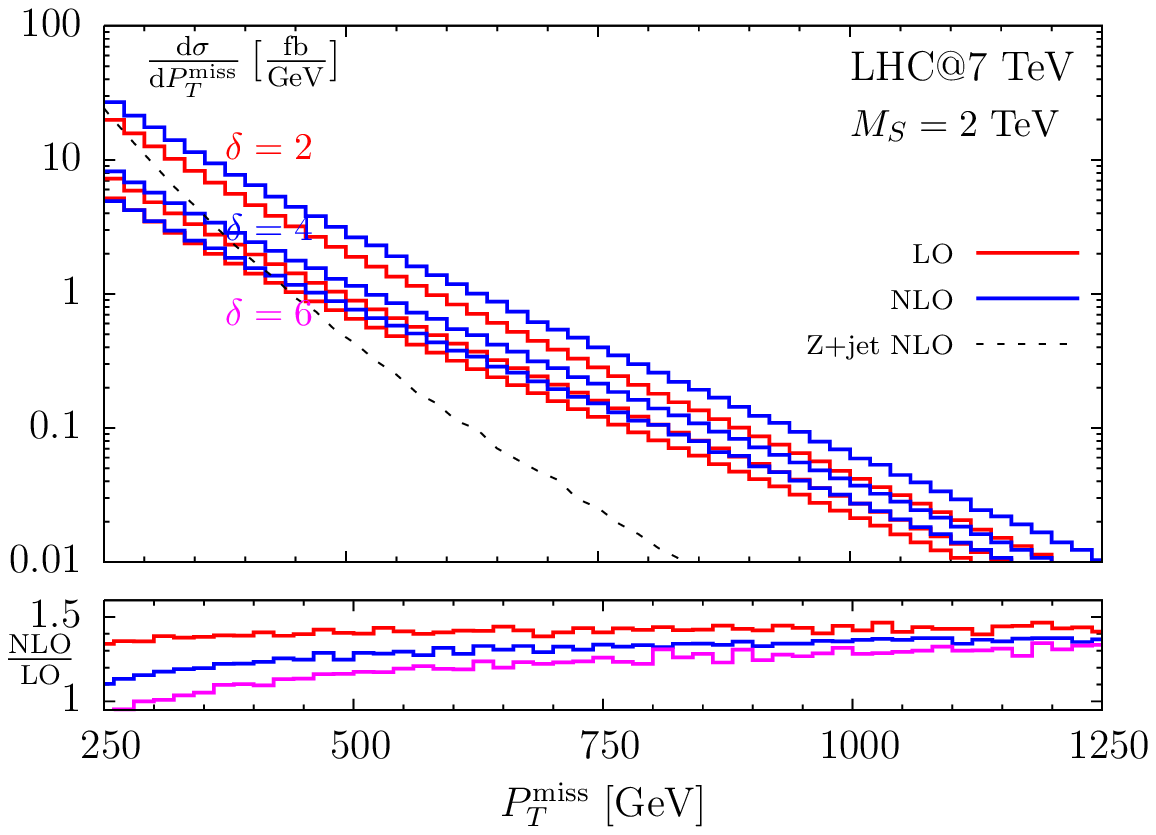}
\caption{$P_T^{\mathrm{miss}}$ distribution for the graviton signal at
  the LHC at 7~TeV cms energy. Also shown is the NLO distribution for the
  dominant $Z\to \nu\bar\nu$ background. The lower part of the plot
  shows $K(P_T)=({\rm d}\sigma_{\rm NLO}/{\rm d}P_T)/({\rm
    d}\sigma_{\rm LO}/{\rm d}P_T)$ for
  $\delta=2,4,6$ (top down).}}

% --------------------------------------------------------------------

\section{Summary}
\label{sec:summary}

We have presented the first calculation of the NLO QCD corrections to
Kaluza-Klein graviton plus jet hadro-production in models with large
extra dimensions. The calculation has been set up as a fully-flexible
parton-level Monte Carlo program\footnote{The fortran code is available upon request from 
karg@physik.rwth-aachen.de.}, and results have been presented for
cross sections and distributions at the Tevatron and at the LHC.%

The QCD corrections stabilize the theoretical prediction and
significantly reduce the scale uncertainty to a level of approximately
10\%. Near the central scale, $\mu = P_T^G$, the QCD corrections
increase the cross section at the LHC by 30-50\%, depending on the
kinematical region and the choice of model parameters. At the
Tevatron, the QCD corrections are modest and negative near $\mu=P_T^G$
and do not strongly depend on the kinematics. A significant
contribution of di-jet events is expected at the LHC, where 20-40\% of
signal events at $P_T^{\mathrm{miss}} > 1$~TeV contain two jets with
$P_T^{j}> 250$~GeV. At the Tevatron, on the other hand, the
contribution of 2-jet events with $P_T^{j}> 60$~GeV is moderate and
does not exceed 20\%. The theoretical uncertainty of the cross section
prediction due to the parton distribution functions is mild, with
approximately 15\% and 5\% at the LHC and the Tevatron, respectively.

We have also studied the uncertainties arising from the UV completion
of the theory by comparing our default NLO results with those
involving a hard truncation scheme. The differences between the two
calculations are small for $\delta=2$ but can reach up to 50\% for
$\delta =6$ and large $P_T$.  Reducing these uncertainties requires to
go beyond the effective field theory approximation of Eq.~(\ref{IL}),
which is beyond the scope of the present paper.  Ignoring form factor
effects for the graviton couplings to gluons and quarks and ignoring
Kaluza-Klein excitations in the loops, defines one particular
phenomenological model. For this model, our calculation quantifies the
size of QCD corrections, and these results may then be taken as an
indication of what to expect of QCD corrections in more complete
models of the UV physics.

%-----------------------------------------------------------------

\acknowledgments 

This work is supported in part by the Deutsche Forschungsgemeinschaft
under SFB/TR-9 ``Computergest\"utzte Theoretische Teilchenphysik'',
the Helmholtz Alliance ``Physics at the Terascale'', and the European
Community's Marie-Curie Research Training Network under contract
MRTN-CT-2006-035505 ``Tools and Precision Calculations for Physics
Discoveries at Colliders''. SK is grateful to Thomas Binoth and
Nikolas Kauer for valuable discussions. SK and MK would like to thank
Arnd Meyer for discussions on experimental aspects of mono-jet
searches.

%-----------------------------------------------------------------

\appendix

\section{Amplitude calculation}
\label{appendixA}
The amplitude for the partonic processes $gg \to Gg$ and $q\bar q
\to Gg$  can be expressed as
\begin{equation}
\begin{aligned}
  \mathcal{M}(g(p_1,\lambda_1) g(p_2,\lambda_2) \to g(p_3,\lambda_3) G(p_4,\lambda_4)) &= \mathcal{M}^{\mu_1\mu_2\mu_3 \mu \nu}   \epsilon_{\mu_1}^{\lambda_1} \epsilon_{\mu_2}^{\lambda_2}  \epsilon_{\mu_3}^{\lambda_3} \epsilon_{\mu \nu}^{\lambda_4}\, f^{abc} \, \\
  \mathcal{M}(q(p_1,\lambda_1) \bar q(p_2,\lambda_2) \to
  g(p_3,\lambda_3) G(p_4,\lambda_4) ) &= \bra{p_2^{\lambda_2}}
  \mathcal{M}^{\mu_3 \mu \nu} \ket{p_1^{\lambda_1}} \,
  \epsilon_{\mu_3}^{\lambda_3} \epsilon_{\mu \nu}^{\lambda_4} \,
  T^a_{ij}\,,
\end{aligned}
\end{equation}
where $\ket{p_i^{\pm}}$ is the Weyl spinor for a massless particle
with momentum $p_i$. Since we consider all quarks to be massless, the
helicity of the quark line is conserved, and therefore we have
$\lambda_1 = \lambda_2 \in [+,-]$.  (Note that the physical helicity
of the anti-quark is given by $-\lambda_2$.)

Applying spinor helicity methods~\cite{Xu:1986xb}, the polarization
vector for a massless spin-1 boson for helicity $\pm$ is given by
\begin{equation}
\epsilon_{\mu}^{\pm}(p,r) = \pm \frac{\bra{r^{\mp} }  \gamma_\mu  \ket{p^{\pm}}  }{\sqrt{2} \braket{r^{\mp} |  {p^{\pm}}}} \, ,
\label{polvector}
\end{equation}
where the vector $r$ in (\ref{polvector}) denotes an arbitrary
reference vector (with $r^2=0)$.

The polarization tensor $ \epsilon_{\mu \nu}^{\lambda_4} $ of the
graviton, a massive spin-2 vector boson, can be constructed from the
polarization vectors of massive spin-1 bosons,
$\epsilon_{\mu}^{\pm,0}$, as follows:\footnote{Note that
  Eq.~(\ref{polvector}) follows the convention $ \epsilon_{\mu
  }^{\lambda,*} = + \epsilon_{\mu }^{-\lambda} $. An additional sign
  is sometimes included in the literature, which would alter the sign
  of the third term in $\epsilon_{\mu \nu}^{0}$ in
  Eq.~(\ref{massivepolvector}) }
\begin{equation}
\begin{aligned}
  \epsilon_{\mu \nu}^{++}  &=   \epsilon_{\mu }^{+} \epsilon_{ \nu}^{+} \\
  \epsilon_{\mu \nu}^{+}  &=   \frac{1}{\sqrt{2} } \left(    \epsilon_{\mu }^{+} \epsilon_{ \nu}^{0} +   \epsilon_{\mu }^{0} \epsilon_{ \nu}^{+}  \right)\\
  \epsilon_{\mu \nu}^{0}  &=   \frac{1}{\sqrt{6} } \left(    \epsilon_{\mu }^{+} \epsilon_{ \nu}^{-} +   \epsilon_{\mu }^{-} \epsilon_{ \nu}^{+}  -2\,   \epsilon_{\mu }^{0} \epsilon_{ \nu}^{0} \right) \\
  \epsilon_{\mu \nu}^{--}  &=   \epsilon_{\mu }^{-} \epsilon_{ \nu}^{-} \\
  \epsilon_{\mu \nu}^{-} &= \frac{1}{\sqrt{2} } \left( \epsilon_{\mu
    }^{-} \epsilon_{ \nu}^{0} + \epsilon_{\mu }^{0} \epsilon_{
      \nu}^{-} \right)\,.
\end{aligned}
\label{massivepolvector}
\end{equation}

In order to extend the spinor-formalism to massive gauge bosons, it is
useful to decompose the graviton momentum into two light-like vectors:
\begin{equation}
p_4 = q_4 + \alpha \, r, \qquad {\textrm{with}} \qquad p_4^2 =m^2, q_4^2=0=r^2, \,\alpha =\frac{m^2}{2\, p_4\cdot r} \, ,
\end{equation}
where the arbitrary reference momentum $r$ can be taken from the list
of available light-like external momenta.  The expressions for the
three polarization vectors $\epsilon^{\pm,0}$ can now be constructed
from the two light-like vectors $p_4$ and $r$ and read
\begin{equation}
\begin{aligned}
\epsilon_{\mu}^{\pm}(p_4,m) &= \pm \frac{\bra{r^{\mp}}  \gamma_\mu  \ket{q_4^{\pm}}  }{\sqrt{2} \braket{r^{\mp}   | p^{\pm}}} \\
\epsilon_{\mu}^{0}(p_4,m) &= \frac{1}{m}\, \left( q_4^\mu - \alpha \, r^\mu  \right) \, .
 \end{aligned}
\label{polvectors}
\end{equation}

Of course, individual helicity amplitudes are no longer independent of
the choice of the reference momentum of the graviton. However, we are
only interested in the spin \emph{sum}, which is independent of the
reference momentum.

By a suitable choice of the reference vectors, we can assemble the
individual spinor products and write them as a trace times a global
spinorial factor. The projector for the $\lambda_1 \lambda_2 \lambda_3
= +++$ helicity combination, for example, reads:
\begin{equation}
  \epsilon_{\mu_1}^{+} \epsilon_{\mu_2}^{+}  \epsilon_{\mu_3}^{+} = \frac{ \spinl{3^-}{\mu_1}{1^-} }{\sqrt{2}\, \Braket{31}}  \frac{ \spinl{1^-}{\mu_2}{2^-} }{\sqrt{2}\,\Braket{12}}  \frac{ \spinl{2^-}{\mu_3}{3^-} }{\sqrt{2}\,\Braket{23}}  = \frac{\tr\left[(1-\gamma_5) 3 \mu_1 1 \mu_2 2 \mu_3\right]}{4\sqrt{2}\, \Braket{31}\Braket{12}\Braket{23} }\, ,
\end{equation}
and the projector for the $\lambda_4 = 2+$ helicity is given by
\begin{equation}
  \epsilon_{\mu \nu}^{++} = \frac{\spinl{1^-}{\mu}{4^-}}{\sqrt{2}\, \Braket{14}} \frac{\bra{4^-} \fmslash{p}_2 \ket{1^-}}{\braket{42} [21]}  \frac{\spinl{1^-}{\nu}{4^-}}{\sqrt{2}\, \Braket{14}} \frac{\bra{4^-} \fmslash{p}_2 \ket{1^-}}{\braket{42} [21]} =  
  \frac{\tr\left[(1-\gamma_5) 1 \mu 4 2 1 \nu  4 2\right]}{4\, \Braket{14}^2 \Braket{42}^2[21]^2 } \, ,
\end{equation}
with the spinor inner products $\braket{ij}= \braket{p_i^- | p_j^+}$,
$[ij]=\braket{p_i^+ | p_j^-}$.

Note that the 40 (20) helicities for $gg\to gG$ ($q\bar q \to gG$) are
related to each other by discrete symmetries, like parity, Bose
symmetry or invariance under charge conjugation. The symmetries
therefore allow for a cross check of the results or can be used to
reduce the algebraical work by calculating only a generic set of
helicity amplitudes. In this way, we could perform the spin sum by
computing only 1 (2) helicity amplitude(s) for $gg\to gG$ ($q\bar q
\to gG$). 

The helicity amplitudes for $gg\to gG$ and $q \bar q \to gG$ summed
over the polarization of the graviton are given by
\begin{equation}
\sum_{\lambda_4} |\mathcal
  M_{gg\to gG}^{+--\lambda_4}|^2 = \frac{2 \hat u^4}{s t u} 
\qquad \mbox{and} \qquad 
\sum_{\lambda_4} |\mathcal
  M_{q \bar q \to gG}^{---\lambda_4}|^2 = \frac{\hat u^2}{2 s t u }
  \left(4 tu +s {m^2_G} \right) 
\label{eq:ggLOhelis}
\end{equation}
where $\hat{u} = u - m_G^2$ etc. Applying Bose and parity transformations 
on the above expressions gives us all helicity amplitudes. 

Summing over final colours and averaging over initial helicities and
colours, the squared LO matrix elements are given by
\begin{equation}
\begin{aligned}
  |\mathcal M_{\text{LO}}|^2(gg\to gG) &=
  \frac{3}{32}\frac{g_s^2}{{\overline M}_{\rm P}^2} \left(4\frac{\hat
      s^4 + \hat t^4 +\hat u^4}{s t u}\right) \, , \\
  |\mathcal M_{\text{LO}}|^2(q\bar q\to gG) &=
  \frac{1}{9}\frac{g_s^2}{{\overline M}_{\rm P}^2} \left( (4 t u+s
    m_G^2) \frac{\hat t^2 + \hat u^2}{stu} \right)\,,
 \end{aligned}
\label{LOmatrixelements}
\end{equation}
in agreement with \cite{feynrules2}, but in a form where the
symmetries are more obvious. The process $qg\to qG$ is related by
crossing to $q \bar q \to gG$.

%%%%%%%%%%%%%% Begin References %%%%%%%%%%%%%%%%%%%%%%%%%%%%%%%%%%%%%%%%

\end{document}